\documentclass{article} 
\usepackage{nips13submit_e,times}
\usepackage{hyperref}
\usepackage{url}
\usepackage{natbib}

\title{Mobile Recommender Systems Methods: An Overview}

\author{
Djallel Bouneffouf \\
Department of Computer Science\\
Telecom SudParis\\
\texttt{Djallel.bouneffouf@it-sudparis.eu}
}
\begin{document}

\maketitle

\begin{abstract}
The information that mobiles can access becomes very wide nowadays, and the user is faced with a dilemma: there is an  unlimited pool of information available to him but he is unable to find the exact information he is looking for. This is why the current research aims to design \textit{Recommender Systems (RS)} able to continually send information that matches the user's interests in order to reduce his navigation time. 
In this paper, we treat the different approaches to recommend.   
\end{abstract}

\section{Introduction}

Recommender System (RS) belong to a more general framework of systems called Personalized Access Systems. These systems integrate the user as an information structure, in the process of selecting relevant information to him \cite{Profile}.

It is possible to classify RS in different ways. The most common classification is based on three  recommendation approaches: Collaborative Filtering, Content-based Filtering and Hybrid-based Filtering \cite{adomavicus, CF, contentfiltering, hybridCF}.

\section{Collaborative Filtering}
The Collaborative Filtering (CF) exploits the user's feedback on resources (documents). The user's assessments are generally represented by some kind of notation, like from one to five starts. They are either assigned explicitly
by users or implicitly from the  indicators of interest.
\\
The first RS has been designated as a CF system by Tapestry \cite{CF} \cite{Breese}. This system allows users access to their emails. Depending on the appreciations of other users of the emails they receive, the system makes recommendations. The authors called this approach "CF" because the users could work together to set undesired emails.
\\
More specifically, CF uses a matrix which rows correspond to the users and the columns to resources. 
Each cell in the matrix corresponds to a note provided by the user for a specific resource. The goal is to predict notes to resources for which the users have not yet provided a rating, and then recommend the best resources w. r. t. the predicted notes. CF is generally classified into two approaches: the memory-based approach and the model-based approach \cite{Breese}. 
\\
- The memory-based approach treats the whole matrix for finding the similarity between users. This approach has the advantage of being both simple to implement and efficient, and allows adapted dynamically when new notes are entered into the matrix, however its complexity is such that its use is only possible in a space data relatively small.
\\
- In the model-based approach, a probabilistic model is built from the users assessments for example (Bayesian classification), and it is applied to find a suitable community of the new user. This model runs faster than the memory-based model, but the construction of the model takes time \cite{Breese}.
\\
\\
CF gives an interesting recommendation only if the overlap between users' history is high (similarity on user's profiles) and the users' content is static. In addition, it suffers from the "cold start" problem since the system can provide relevant recommendations only if the user provides enough rating for a sufficient number of resources.

\section{Content-Based Filtering}
The Content-Based Filtering (CBF) analyses the resources  to determine what resources are likely to be interesting for a given user. This domain is highly similar to IR. Indeed, the same techniques are used, the difference being essentially in the absence of explicit requests made by the user. Therefore, many of the general concepts of recommendation based on the content come from information retrieval \cite{contentfiltering}.
\\
CBF recommendation is based on notes assigned by users to multiple documents and the similarity between documents according to certain criteria. For example, in order to recommend websites not previously seen to a user, the system searches for similarities with respect to certain characteristics (type of website, content, etc.), with websites that have been previously assessed a high score by the same user. From the similarity computation, only websites with a high degree of similarity will be recommended \cite{BS97, Pazzani, adomavicus}.
\\
Generally, the content of a document is described by keywords; and for this, the system must know the importance of each word in a document associating it with a weight. Weights can be computed in several ways, being the most known the so called TF/IDF method. This method tries to find the importance of a word in a text using its "term frequency" (TF) and "inverse document frequency" (IDF) values. In IR, terms are the resulting word roots after a pre-processing step including steeming and eliminating stop-words. .

Let \textit{N} be the number of documents in the collection $D$ and $t_i$ a term to recommend.
\\
Let $f_{i,j}$ be the number of times $t_i$ appears in document $d_j \in D$. $f_{z, j}$ represents the frequency of term $t_z \neq t_i$ in $d_j$. Thus the relative frequency $TF_ij$ of term $t_i$  in document $d_j$ is defined by Eq.~\ref{eq:tfidf}:
\begin{equation}
\label{eq:tfidf} 
TF_{i,j} = f_{i,j}/max_z f_{zj}
\end{equation}
The $IDF_i$ value of term $t_i$ in document $d_j$ is defined by Eq.~\ref{eq:idfi}:
\begin{equation}
\label{eq:idfi} 
IDF_i =Log(N/n_i)
\end{equation}
Then the TF-IDF weight of term $t_i$ in document $d_j$ is given by Eq.~\ref{eq:wij}:
\begin{equation}
\label{eq:wij} 
w_{i, j} = TF_{i, j} x IDF_{i}
\end{equation}
In Eq.~\ref{eq:wij} the content of a document is defined by \textit{Content}$(d_j) = (w_{1J}, ..., w_{KJ})$.
\\
CBF identifies new documents which match an existing user's profile. However, the recommended documents are always similar to the documents previously selected by the user \cite{15}. The main limitation of content-based recommendation is that it requires acquiring a sufficient number of attributes that describes the resources. That is why it is appropriate in the context of text resources or when textual descriptions of resources have been entered manually.  
\\
Another limitation is that a new user of such a system must provide relevance assessments for a minimum number of resources before the system can provide relevant recommendations (cold start problem). 

\section{Hybrid Approach}
This approach combines the collaborative and the content-based filtering methods to reduce their limitations. Among the several ways to combine these two methods, we distinguish the following ones. 
\subsection{Extending the CBF with properties of CF} Several hybrid recommendation systems use this method, as \cite{BS97, PAZ}. In these works, authors compute the similarity between two users profiles based on all the documents each user assessed, rather than only on assessed documents users have in common. The advantages of this method is to decrease the cold start problem if the users have not assessed the same documents. 
\subsection{Extending the CF with the properties of CBF} these techniques are used to make clustering on a group of profiles based on the content. For example, in \cite{SN99}, authors use latent semantic indexing to create a set of user's profiles.
\subsection{Unifying CBF approach and CF} An example of unification is the work described in \cite{SN99}, where authors propose to use characteristics of both methods in a single classifier, that use Bayesian regression to classify the interring and the non-interesting documents.
\\
Studies in \cite{SN99}\cite{PAZ}\cite{baeza1999} have shown that the hybrid approach has given better results compared to pure recommendation approaches like content or CF. 

\section{Discussion}
We present now a synthesis of the recommendation approaches.
\\
The approaches are grouped in Table~\ref{tab:recommendationapproaches} according to the  recommendation approaches and their advantages / disadvantages. 
\begin{table} [h]
\caption{ The recommendation approaches}
\label{tab:recommendationapproaches}       
\begin{tabular}{|p{5cm}|p{4cm}|p{4cm}|p{3cm}|}
\hline
\bf Recommendation approach         & \bf Advantage & \bf Disadvantage  
\\
\hline Collaborative Filtering	\cite{CF} \cite{Breese}    & Gives an interesting recommendation when the overlap between users' history is high and the users' content is static  &  inability to recommend new resources 
\\
\hline Content-Based Filtering \cite{BS97, Pazzani, adomavicus}  & Recommends new documents to users & Can not make recommendation when a new user arrive  
\\
\hline Hybrid Approach \cite{SN99}\cite{PAZ}\cite{baeza1999}  & Overcome the limitations of the both approaches ( can make recommendation when a new user arrive and recommend new resources) & Does not follow the evolution of the user's contents
\\
\hline
\end{tabular}
\end{table}

From the Table~\ref{tab:recommendationapproaches}, we can observe that the inability of CF to recommend new documents is reduced by combining it with the CBF technique \cite{13}. However, the user's content in a real world undergoes frequent changes. These issue makes content-based and CF approaches difficult to apply \cite{8}.

\bibliographystyle{named}
\bibliography{profile}

\end{document}